\documentclass[preprint,showpacs,preprintnumbers,amsmath,amssymb, showkeys]{revtex4}

\usepackage{graphicx}
\usepackage{dcolumn}
\usepackage{pstcol,times,color,graphicx,graphics,pstricks,pst-node,pst-coil,pst-grad, pst-plot, pst-3d, pst-fill, multido, pst-poly, pst-tree, url}
\usepackage{bm}
\newpsobject{showgrid}{psgrid}{subgriddiv=1,griddots=10,gridlabels=6pt}

\begin{document}

\preprint{APS/123-QED}

\title{The determination of the apsidal angles and  Bertrand's theorem}

\author{Filadelfo C. Santos}
 \email{filadelf@if.ufrj.br}

\author{Vitorvani Soares}%
 \email{vsoares@if.ufrj.br}

 \author{Alexandre C. Tort}
 \email{tort@if.ufrj.br}
\affiliation{Instituto de F\'{\i}sica -- Universidade Federal do Rio de Janeiro \\
Cidade Universit\'aria, P.O.B. 21945-970, Rio de Janeiro, Brazil}

\date{\today}
\begin{abstract}
We derive an expression for the determination of the apsidal angles that holds good for arbitrary central potentials. Then we discuss under what conditions the apsidal angles remain independent of the mechanical energy and angular momentum in the central force problem.  As a consequence, an alternative and non-perturbative proof of Bertrand's theorem is obtained.
\end{abstract}

\maketitle
\section{Introduction}
In 1873, J. Bertrand\cite{bertrand} published a short but important
paper in which he proved that there are only two central fields for
which all orbits radially bounded are closed, namely: The isotropic harmonic oscillator field, 
and the gravitational one. Because of this
additional degeneracy -- in the language of modern group theory we associated the unitary group   $U(3)$  with the harmonic oscillator 
and the orthogonal group $O(4)$ with the gravitational potential --  it is no wonder that the properties of
those two fields have been under close scrutiny since Newton's
times. Newton addresses to the isotropic harmonic oscillator in
proposition X Book I of the \textit{Principia}, and to the inverse-square
law in proposition XI \cite{Newton}. Newton shows that both fields give rise to
elliptical orbits with the difference that in the first case the
force is directed toward the geometrical center of the ellipse and
in the second case the force is directed to one of the foci.  Bertrand's proof is concise, 
elegant, and,  contrary to what one may be led to think by a number of
perturbative demonstrations that can be found in 
textbooks and papers on the subject, fully non-perturbative. As
examples of perturbative demonstrations the reader can consult
references \cite{tikochinsky, Brown, Zarmi}. We can also find in the
literature demonstrations that resemble the spirit of Bertrand's
original work as for example \cite{greenberg, Arnold}. 
All perturbative demonstrations and most of the non-perturbative ones, however, have a 
restrictive feature, to wit, they set a limit on the number of
possibilities of the existence of central fields with the property
mentioned above to a finite number and finally show explicitly that
among the surviving possibilities only two, the newtonian and the
isotropic harmonic oscillator, are  really possible.

Let us now outline briefly Bertrand's approach to the problem. In his paper, Bertrand initially proves by taking into consideration
the equal radii limit that a central force $f(r)$ acting on a
point-like body able of generating radially bounded orbits must
necessarily be of the form

\[ f\left( r\right) =\kappa\,r^{\left(1/p^2-3\right)} ,\]
where $r$ is the radial distance to the center of force, $\kappa$ is a
constant and  $p$ a rational number. Next, making use of this
particular form of the law of force and considering also an
additional limiting condition, Bertrand finally shows that only for
$p=1$ and $p=1/2$, which correspond to Newton's gravitational law of
force

\[f\left( r\right) =-\frac{\kappa}{r^{2}} ,\]
and to the isotropic harmonic oscillator law of force

\[f\left( r\right) = - \kappa\,r,\]
respectively, we can have orbits with the properties stated in the theorem.
Moreover, Bertrand can also prove that only for these laws of force all bounded orbits are closed.
For more details the reader is referred  to the original paper or to its English translation  \cite{bertrand}.

In the present paper, we consider the issue from an alternative point of view. This alternative approach stems from a different kind of question that we can ask ourselves. For radially bounded orbits there are two extreme values for the possible radii, namely, a maximum and a minimum one. These extreme values were named by Newton himself superior apse and inferior apse, respectively, or briefly, apses. The angular displacement between these two successive points or apses defines the apsidal angle $\Delta\theta_a$. In a central field of force, the apsidal angle depends on the total mechanical energy $E$ and the magnitude of the angular momentum $\ell$ . The question we can formulate is the following: \textit{For what potentials the apsidal angle has the same value for all orbits -- that is for arbitrary energy and angular momentum -- and what is its value?} The answer as we will show in the next sections is that the apsidal angle is constant for only two potentials: The Newtonian and the isotropic harmonic oscillator one, and consequently as a corollary we have an alternative and  non-perturbative proof  of Bertrand's theorem.
\section{The apsidal angle}

In a central field, in which the magnitude of the force $\mathbf{f} (\mathbf{r}) $ depends only on the distance $r=\|\mathbf{r}\|$ to the center of force, we can introduce the potential function $V( r) $ with the property $\text{\boldmath
$f$}( \mathbf{r}) =-\mathbf{\nabla} V({ r}) $ such that the total mechanical energy $E$ of a particle with mass $m$ orbiting in this field is a constant of motion.
Thus we write
\begin{align}
E&=\frac{m}{2}{\text{\boldmath $v$}}^{2}+V(r) ,\ \ \
\end{align}
\noindent where $\text{\boldmath $v$}$ is the velocity of the particle. Moreover, the conservation of the angular momentum $\text{\boldmath $l$}$ of the particle in a central field constrains its motion to a fixed plane and allows for the introduction of an effective potential defined by
\begin{equation}
U( r) =V( r) +\frac{l^{2}}{2m r^{2}} ,%
\label{XRef-Equation-82110330}
\end{equation}
\noindent with the help of which it is possible to reduce the motion to an equivalent unidimensional problem. Let us define for convenience the variable $z=1/r$ and the functions $v(z) =V(1/z)$ and $u(z)=U(1/z)$ in such a way that the effective potential (\ref{XRef-Equation-82110330}) is now written
\begin{equation}
u( z) =v( z) +\frac{l^{2}}{2m}z^{2} . %
\label{XRef-Equation-82110427}
\end{equation}
Let us also assume that the effective potential (\ref{XRef-Equation-82110427})
has a minimum at $z_{0}=1/r_{0}$, such that the particle can move in a circular orbit with radius $r_{0}$.  This particular value of $z$ is determined by the condition
\begin{equation}
u^{\prime }( z_{0}) =v^{\prime }( z_{0}) +\frac{l^{2}}{m}z_{0}=0,
\label{XRef-Equation-821105340}
\end{equation}
from which we obtain

\begin{equation}
l^{2}=-{mv}^{\prime }( z_{0}) /z_{0}) . 
\end{equation}
Taking this result into the effective potential ( \ref{XRef-Equation-82110427}) we can write

\begin{equation}
u(z) =v(z) -\frac{v^{\prime }( z_{0}) }{2z_{0}}z^{2}.
\label{uu}
\end{equation}
In this way we see that the first derivative of the effective potential at the point $z=z_0$ is zero for each and every potential function $v(z)$. This means that the vanishing of this derivative does not impose restrictions of any kind on the potential function $v(z)$. The second derivative at the point $z=z_0$ is  

\begin{equation}
u^{\prime\,\prime }( z_{0}) =v^{\prime\,\prime }( z_{0}) - \frac{v^{\,\prime} (z_0)}{z_0} .
\end{equation}
If $u^{\prime\,\prime }( z_{0})$ is zero for an arbitrary point $z_0$, we see that the potential must be of the form
$v(z)=a+b\,z^2$, and in this case the force must be of the form $\kappa/r^3$. Newton in the Proposition IX of Book I of the Principia shows that in this case the orbit is an equiangular spiral and therefore without apsidal points. Hence, in what follows we will suppose that the second derivative of the effective potential is not null. 

If now we assume that $u$ and $z$ are complex variables and perform an analytical continuation of the function  $ u(z) $ we can determine the inverse function $z=z(u) $ by applying the generalization of the B\"urmann-Lagrange series for a multivalued inverse function \cite{markuanalytical}.  With this purpose in mind we choose the point $u_0=u(z_0)$  as the point around which we will perform the expansion of this function. The result is
%
\begin{equation}
z=z_0+\sum_{n=1}^\infty\,\frac{1}{n!}\,\left[ \frac{d^{n-1}}{d\zeta^{n-1}}\,\chi^n\,\left( \zeta\right) \right]_{\zeta=z_0}\,\left(u-u_0\right)^{n/2} ,
\end{equation}

\noindent where
\begin{equation}
\chi ( \zeta ) =\frac{\zeta -z_{0}}{{\left[ u( \zeta ) -u( z_{0}) 
\right] }^{\frac{1}{2}}} ,
\label{chi}
\end{equation}
and  $u_0=u(z_{0}) $ is a first order algebraic branching point of the function $z(u)$. It follows that we can define two real inverse functions of the function $ u=u(z) $; the first one is
\begin{equation}
z_1=z_{0}+\sum \limits_{n=1}^{\infty }{C_{n}( u-u_{0}) }^{\frac{n}{2}},%
\label{XRef-Equation-821154818}
\end{equation}
\noindent which holds for $z>z_{0} $ (or $r<r_{0}$). The second one is
\begin{equation}
z_2=z_{0}+\sum \limits_{n=1}^{\infty }{\left( -1\right) }^{n}{C_{n}(
u-u_{0}) }^{\frac{n}{2}},%
\label{XRef-Equation-821154910}
\end{equation}
\noindent which holds for $z<z_{0} $ (or $r>r_{0}$), and where we have written for convenience
\begin{equation}
{\left. C_{n}=\displaystyle{\frac{1}{n!}}\frac{d^{n-1}\chi ^{n}( \zeta ) }{{\mathrm{d\zeta
}}^{n-1}}\right| }_{\begin{array}{l}\zeta =z_{0}
\end{array}}.
\label{Cn}
\end{equation}
It is also convenient to write the function (\ref{chi}) in the form
\begin{equation}
\chi ( \zeta ) =\frac{1}{\sqrt{ \displaystyle{ \frac{v( \zeta ) -v( z_{0}) -v^{\prime
}( z_{0}) \left( \zeta -z_{0}\right) }{{\left( \zeta -z_{0}\right)
}^{2}}-\frac{v^{\prime }( z_{0}) }{2z_{0}}} } },
\label{chi 2}
\end{equation}
which can be obtained by making use of (\ref{uu} ). For radially bounded orbits the two extreme values for the radii, namely, $r_{\max
}$ and $r_{\min }$, maximum and minimum  (or $z_{\min
}$ and $z_{\max }$), respectively are determined by the condition $\dot{r}_{a}=0$, or $\dot{z}_a=0$. 
The particle oscillates indefinitely between $r_{\max
}$ and $r_{\min }$. In terms of the effective potential radially closed orbits are characterized 
by extreme points that satisfy the condition
$E=U( r_{a}) $, or $E=u(z_a)$. For convenience we take the direction defined by the arbitrary vector ${\text{\boldmath $r$}}_{0}$ as the reference for the measure of angular displacements.  Hence, the angular displacement between two successive apses.  that is, the apsidal angle $\Delta\,\theta_{a}$, can be written in the form
\begin{equation}
\Delta \theta_{a}=\Delta\,\theta_{1}+\Delta\,\theta_{2},
\label{Theapsidalangle}
\end{equation}
\noindent where $\Delta\,\theta_{1}$ is the angular displacement from the point ${\text{\boldmath $r$}}_{0}$ to the point ${\text{\boldmath $r$}}_{\min }$ and $\Delta\,\theta_{2}$ is the angular displacement to the apsidal point ${\text{\boldmath
$r$}}_{\max }$.

The angular displacement in a central field of force can be easily determined from the conservation laws of the mechanical energy and the angular momentum. Hence we can write
\begin{equation}
\Delta \theta_{1}=-\int _{r_{0}}^{r_{\mathit{\min }}} \displaystyle{\frac{\ell}{m r^{2}}} \frac{dr}{\sqrt{\displaystyle{\frac{2}{m}}\left(
E-U\right) }},
\end{equation}
\noindent and
\begin{equation}
\Delta \theta_{2}=\int _{r_{0}}^{r_{\max }}\frac{\ell}{m r^{2}}\frac{d r}{\sqrt{\displaystyle{\frac{2}{m}}\left(
E-U\right) }}.
\end{equation}
\noindent Making use of the inverse functions (\ref{XRef-Equation-821154818}) and (\ref{XRef-Equation-821154910}) we have
\begin{eqnarray}
\Delta \theta_{1}&=&\frac{\ell}{\sqrt{2m}}\int _{U_{0}}^{E}\frac{{dz}_{1}}{dU}\frac{dU}{\sqrt{\left(
E-U\right) }} \nonumber \\ &=& \frac{\ell}{2\sqrt{2m}}\sum \limits_{n=1}^{\infty
}{{n\,C}}_{n}\int _{\begin{array}{l}
  U_{0}
\end{array}}^{E}\frac{{\left( U-U_{0}\right) }^{\frac{n}{2}-1}}{\sqrt{E-U}}\,d
U ,%
\label{XRef-Equation-82214290}
\end{eqnarray}
and
\begin{eqnarray}
\Delta \theta_{2}&=&\frac{\ell}{\sqrt{2m}}\int _{U_{0}}^{E}\frac{{{dz}}_{2}}{{dU}}\frac{{dU}}{\sqrt{\left(
E-U\right) }} \nonumber \\ &=&\frac{\ell}{2\sqrt{2m}}\sum \limits_{n=1}^{\infty
}n {\left( -1\right) }^{n }\,C_{n}\int _{\begin{array}{l}
 U_{0}
\end{array}}^{E}\frac{{\left( U-U_{0}\right) }^{\frac{n}{2}-1}}{\sqrt{E-U}}\,d
U .%
\label{XRef-Equation-822142931}
\end{eqnarray}
The integrals in equations (\ref{XRef-Equation-82214290}) and (\ref{XRef-Equation-822142931}) can be evaluated with the help of 

$$ \int_a^b\,\left( x-a \right)^{\lambda-1}\left( b-x \right)^{\nu-1} \,dx = \left(b-a\right)^{\lambda+\nu-1}\,B\left(\lambda, \nu \right) , $$
where

$$B\left(\lambda, \nu\right)=\frac{\Gamma (\lambda) \Gamma (\nu)}{\Gamma (\lambda + \nu) } ,$$
which holds for $b > a$, $\Re\, \mu > 0$ and $\Re\, \nu > 0$, see \cite{Grad}, Formula 3.196.3.  Therefore we have
\begin{eqnarray}
\Delta \theta_{1} &=&
\frac{\ell\,\pi}{2\sqrt{2m}}\sum \limits_{k=0}^{\infty }\left(
2k+1\right) C_{2k+1}\frac{\left( 2k-1\right) !!}{\left( 2k\right)
!!}{\left( E-U_{0}\right) }^{k}\nonumber \\ &+& \frac{\ell}{2\sqrt{2m}}\sum
\limits_{k=0}^{\infty }2(k+1)C_{2k+2}\frac{2\left( 2k\right)
!!}{\left( 2k+1\right) !!}{\left( E-U_{0}\right) }^{k+\frac{1}{2}} ,
\label{XRef-Equation-822152230}
\end{eqnarray}
\noindent and
\begin{eqnarray}
{\Delta \theta }_{2}&=&
\frac{\ell \pi}{2\sqrt{2m}}\sum \limits_{k=0}^{\infty }\left( 2k+1\right)
C_{2k+1}\frac{\left( 2k-1\right) !!}{\left( 2k\right) !!}{\left(
E-U_{0}\right) }^{k}\nonumber \\ &-& \frac{\ell}{2\sqrt{2m}}\sum \limits_{k=0}^{\infty
}2(k+1){C}_{2k+2}\frac{2\left( 2k\right) !!}{\left( 2k+1\right) !!}{\left(
E-U_{0}\right) }^{k+\frac{1}{2}} .
\label{XRef-Equation-822152316}
\end{eqnarray}

\noindent Notice that the first term in the first summation is the only one that does not depend on the energy. 
 Notice also that the constants $C_{n}$ do not depend on the energy as well. After adding equations (\ref{XRef-Equation-822152230})
and (\ref{XRef-Equation-822152316}) we obtain for the apsidal angle the expression
\begin{equation}
\Delta \theta_{a}=\frac{\ell \pi}{\sqrt{2m}}\sum \limits_{k=0}^{\infty }\left( 2k+1\right)
C_{2k+1}\frac{\left( 2k-1\right) !!}{\left( 2k\right) !!}{\left(
E-U_{0}\right) }^{k} .
\label{apsidal angle}
\end{equation}
\noindent Notice that this series does not depend on the even coefficients. Equations (\ref{XRef-Equation-822152230}), (\ref{XRef-Equation-822152316}), and (\ref{apsidal angle}) are our main results and in the next section we will use them to answer the question we posed at the beginning of this work, to wit:  For what potentials the apsidal angle has the same value for all orbits, i.e.: for arbitrary  values of the energy and angular momentum, and what value does it assumes. 

\section{The Newtonian potential}

Let us begin by looking for potentials that keep the partial angles $\Delta\,\theta_1$ and $\Delta\,\theta_2$ separately constant. If this happens to be so, the apsidal angle $\Delta\,\theta_a$ will also be constant for any value of the energy and the angular momentum.  
 In order to accomplish that it is necessary that for $n\geqslant 2$ all coefficients  $C_{n}$
in equations (17) and (18) be  equal to zero.  After evaluating the coefficient $C_{2}$  and setting it equal to zero we obtain $v^{\,\prime\prime\prime}(z_0)=0$, and taking into account that $z_0$ is an arbitrary point it follows that
\begin{equation}
v^{\prime \prime \prime }( z) =0.
\label{vdiff}
\end{equation}
The general solution of  (\ref{vdiff}) is
\begin{equation}
v(z) =a z^{2}+b z+c .
\label{XRef-Equation-822154852}
\end{equation}
\noindent The integration constant $c$ is an additive term to the potential and can be discarded without loss of generality.  
The second derivative of the effective potential at the stationary point is $u^{\,\prime\,\prime}=-b/z_0$. It follows that $z_0$ is effectively a minimum 
only if the constant $b$ is negative and this corresponds to an attractive field. For a potential function of the form given by (\ref{XRef-Equation-822154852}) we can evaluate the function $\chi (\zeta ) $ given by  (\ref{chi 2}) to obtain
\begin{equation}
\chi (\zeta ) =\frac{1}{\sqrt{-\displaystyle{\frac{b}{2z_{_{0}}}}}} .
\end{equation}

\noindent  Hence the only non-zero coefficient is $C_{1}$, because the function $\chi (\zeta ) $, for this potential, does not depend on $\zeta$ and therefore $\Delta \theta_{1}=$ $\Delta\,\theta_{2}=\mbox{constant}$.
Evaluating the first coefficient for this potential we obtain
\begin{equation}
\Delta \theta_{1}=\Delta \theta_{2} =\frac{\pi }{2}\sqrt{1+\displaystyle{\frac{2\,a\, z_{{0}}}{b}}} .%
\label{XRef-Equation-822161234}
\end{equation}

For a fixed value of the angular momentum any potential of the form given by (\ref{XRef-Equation-822154852})  generates bounded orbits with equal and constant angles $\Delta\,\theta_1$ and $\Delta\,\theta_2$ and thus the apsidal angle $\Delta\,\theta_a$ as defined by  (\ref{Theapsidalangle})   will also be constant.  Moreover,  if only the constant $a$ vanishes, that is, if  the potential is of the form $v(z)=b\,z$,  it is possible to obtain a potential for which the apsidal angle is also independent of the angular momentum.  
This potential corresponds to the newtonian one,  the  apsidal angle is $\pi $ and the orbits are closed. The vector $\mathbf{r}_{0} $ is perpendicular to the vectors $\mathbf{r}_{\max }$  and
$\mathbf{r}_{\min }$, which in their turn are antiparallel vectors with respect to each other
defining in this way only one symmetry axis of the trajectory of the
particle.

\section{The isotropic harmonic oscillator potential}

\noindent Let us now look for potentials that keep the apsidal angle $\Delta \theta_a$ 
constant. In order to accomplish this it is necessary that all the odd
coefficients $C_n$ starting from the third one be zero. As before we begin
by explicitly calculating the lowest order coefficient, namely
$C_3$. Evaluating this coefficient, setting it equal to zero, and considering the arbitrariness of $z_0$ we
obtain the following differential equation
\begin{equation}
\frac{5}{3}{v^{\,\prime \prime \prime }}^{2}( z) -\left[ v^{{\,\prime\prime}}(
z) -\frac{v^{\,\prime}( z) }{z}\right] v^{\left( iv \right) }(
z) =0 .
\label{masterdiffeq1}
\end{equation}
\noindent Writing the function $v(z)$ as
\begin{equation}
v(z) =\frac{z^{2}}{2}\int \frac{\phi ( z) }{z}dz-\frac{1}{2}\int
z \varphi ( z) dz ,
\label{integralsolution}
\end{equation}

\noindent we can recast (\ref{masterdiffeq1}) into the form
\begin{equation}
\frac{5}{3}\frac{1}{z \varphi ( z) }\frac{d}{d z}[ z \varphi ( z)
] -\frac{\frac{d}{d z}[ \frac{1}{z}\frac{d}{d z}[ z \varphi ( z)
] ] }{\frac{1}{z}\frac{d}{d z}[ z \varphi ( z) ] }=0.
\label{masterdiffeq2}
\end{equation}
Equation (\ref{masterdiffeq2}) can be immediately integrated and its general solution is

\begin{equation}
\phi (z)=\frac{1}{z}\left( Az^2+B \right)^{-3/2},
\label{mastersolution}
\end{equation}
where $A$ and $B$ are arbitrary constants. Notice now that the newtonian potential analyzed before, which is a solution of (\ref{vdiff}),  is also necessarily solution of (\ref{masterdiffeq1}) as can be immediately verified.  Such a solution can be obtained from  (\ref{mastersolution}) by setting $A=0$. Therefore, without any loss of generality we suppose $A \neq 0$ and for convenience recast (\ref{mastersolution}) into the form

\begin{equation}
\phi (z) = \frac{4\,k}{z}\left(z^2+b^2\right)^{-3/2},
\label{fi}
\end{equation}
where $k \neq 0$ and $b$ are new constants. We write $b^2$ in order to assure that the function $\phi (z)$, and consequently 
the corresponding potential function, does not become ill-defined in the region around $z=0$.  
Making use of (\ref{fi}) and (\ref{integralsolution}) we can determine the potential function $v(z)$.  For $b= 0$ we  obtain

\begin{equation}
v (z) =\frac{k}{2z^{2}}+c z^{2}+d,
\label{sol1}
\end{equation}
and for $b\neq 0$

\begin{equation}
v (z) = - \frac{4kz\sqrt{z^2+b^2} }{b^4}+c z^{2}+d.
\label{sol2}
\end{equation}
The constant $d$ is a simple shift of the zero of the potential and does not influence the law of force.
Considering first the potential given by  (\ref{sol1}) and evaluating the second derivative of the effective potential at the stationary point we obtain

\begin{equation}
u^{\,\prime\prime} \left( z_0 \right) =  \frac{{2k}}{{z_0^4 }}.
\end{equation}
It follows that we will have a minimum only if $k > 0$. Making use of (\ref{chi 2})  we have

\begin{equation}
\chi \left( \zeta \right) = \sqrt {\frac{2}{k}}\, \frac{{z_0^2\, \zeta}}{{\zeta + z_0}}. 
\label{chi 3}
\end{equation}
Performing the analytical continuation of (\ref{chi 3})  and making use of the Cauchy formula we obtain from (\ref{Cn})

\begin{equation}
C_n  = {\displaystyle{1 \over {n}}}\left( {\frac{2}{k}} \right)^{\frac{n}{2}}\,\frac{1}{2\pi i} \,z_0^{2n}  \oint_{C_{z_0}}\,
{\frac{{\zeta^n }}{{\left( {\zeta^2  - z_0^2 } \right)^n }}}\, d\zeta .
\label{Cn1}
\end{equation}
The integral in (\ref{Cn1} ) can be easily evaluated by making the transformation $\eta = \zeta^2  - z_0^2 $ in the neighborhood of $z_0$. It follows then

\begin{equation}
C_n  = {\displaystyle{1 \over {n}}}\left( {\frac{2}{k}} \right)^{\frac{n}{2}}\,\frac{1}{2\pi i} \, z_0^{2n}  \oint_{C_{z_0}}\, 
{\frac{{\left( {\eta + z_0^2 } \right)}}{{\eta^n }}^{\frac{{n - 1}}{2}} }\,d\eta .
\label{Cn2}
\end{equation}
The residue of (\ref{Cn2}) is obviously null if $n$ is an odd integer greater than one. Therefore, we can be sure that all potentials given by (\ref{sol1}) generate orbits for which the apsidal angles depend only on the angular momentum.  The apsidal angle for this class of potentials follows from (\ref{apsidal angle}) and is given by

\begin{equation}
\Delta \theta _a  = \frac{{l\pi }}{{\sqrt {2m} }}C_1  = \frac{\pi }{2}\sqrt {1 - \frac{{cz_0^4 }}{{2k}}}  .
\label{apsidal angle 2}
\end{equation}
In order to have an apsidal angle not dependent on the angular momentum it is necessary that the constant $c$ be set equal to zero. This will lead us to potentials of the form
 
\begin{equation}
v\left( z \right) = \frac{k}{{2z^2 }},
\end{equation}
which corresponds to the isotropic harmonic oscillator. In this case (\ref{apsidal angle 2}) yields
\begin{equation}
\Delta \theta _a  = \frac{\pi }{2},
\end{equation}
and,  therefore, the orbit is  closed.

\section{The additional third potential}

Finally let us consider the potential given by (\ref{sol2}) for which the effective potential reads

\begin{equation}
u\left( z \right) =  - \frac{{a\,z\,\sqrt {z^2  + b^2 } }}{{b^4 }} + \frac{{a\left( {4z_0^3  + 2b^2 z_0} \right)\,z^2 }}{{4b^4 z_0^2 \sqrt {z_0^2  + b^2 } }}. 
\label{uz}
\end{equation}
The second derivative of $u(z)$ at $z_0$ is easily evaluated and the result is

\begin{equation}
u^{\,\prime\prime} \left( z_0 \right) = \frac{a}{{z_0\sqrt {\left( {z_0^2  + b^2 } \right)^3 } }}.
\end{equation}
We can see that only for $a > 0$  we will have a minimum and in what follows will show that this must indeed be the case here. Making use of (\ref{chi 2})  we obtain after some algebraic manipulations

\begin{equation}
\chi \left( \zeta \right) = \sqrt {\frac{1}{{2k}}} \sqrt {\sqrt {z_0^4  + b^2 z_0^2 } } \; \; \frac{ \sqrt{ \left(b^2+2z_0^2 \right)  \zeta^2+2\sqrt{z_0^4+b^2z_0^2}\sqrt{\zeta^4+b^2\zeta^2}  +b^2z_0^2} } {\zeta + z_0} .
\label{cf1}
\end{equation}
Next we consider the analytical continuation of this function. 
The complex function given by (\ref{cf1})  has algebraic branching points at $z=\pm\,z_0$ and $z=\pm\,ib$ and is defined on a four-sheeted Riemann surface.  Hence we  choose a sheet on this Riemann surface such that the functions $\chi (\zeta)$ does not have branch points at $z=\pm z_0$ and with a branch cut along the imaginary axis from $z=ib$ to $z=-ib$. Notice that in this case the numerator of  (\ref{cf1})  does not have  zeroes.  Making use of Cauchy formula we can write

\begin{equation}
\left.{\frac{{d^{n - 1} \chi ^n \left( z \right)}}{{dz^{n - 1} }}}\right|_{z=z_0}  = \frac{(n-1)!}{2\pi i}\,\oint\limits_{C_{z_0} } \, \vartheta^n\left(\zeta\right)\,d\zeta ,
\label{cfd1}
\end{equation}
where we have defined

\begin{equation}
\vartheta \left(\zeta\right) =\sqrt {\frac{1}{{2k}}} \frac{\sqrt{z_0^4+b^2z_0^2}\sqrt{ \left(b^2+2z_0^2 \right)  \zeta^2+2\sqrt{z_0^4+b^2z_0^2}\sqrt{\zeta^4+b^2\zeta^2}  +b^2z_0^2} }{\zeta^2-z_0^2}
\label{cfd2}
\end{equation}
and $C_{z_0}$ is a small closed path encircling the point $z_0$. The integrand in equation (\ref{cfd1}) has another pole at the point $z=-z_0$. Then, choosing the path indicated in Figure (\ref{contour})  the following results can be easily obtained

\begin{equation}
\mathop {\lim }\limits_{R \to \infty} \oint\limits_{C_R }  \vartheta^n (\zeta) d\zeta = 0 ,
\end{equation}
if $n > 1$;

\begin{equation}
\mathop {\lim }\limits_{r_2  \to 0} \oint\limits_{C }\vartheta^n (\zeta) d\zeta = 0 ,
\end{equation}
and
\begin{equation}
\mathop {\lim }\limits_{r_4  \to 0} \oint\limits_{C } \vartheta^n (\zeta) d\zeta = 0.
\end{equation}
Finally, the integral over the finite parts of the real axis and the imaginary axis, see Figure (\ref{contour}),  cancel out. Then we have

\begin{equation}
\int \limits_{\zeta=z_0 } \vartheta^n (\zeta) d\zeta + \int \limits_{\zeta=-z_0 } \vartheta^n (\zeta) d\zeta=0,
\end{equation}
for $n > 1$. We can easily prove by making the substitution $\zeta \to -\zeta $ in the second integral above that
\begin{equation}
\left[ {1 + \left( { - 1} \right)^{n + 1} } \right] \int \limits_{\zeta=z_0 } \vartheta^n (\zeta) d\zeta = 0 ,
\end{equation}
and therefore we conclude that

\begin{equation}
\left. {\frac{{d^{n - 1} \chi ^n \left( z \right)}}{{dz^{n - 1} }}} \right|_{z = z_0}  = 0 ,
\end{equation}
if $n$ is an odd integer greater than one. The last step is the evaluation of the apsidal angle for this case. Making use of (\ref{XRef-Equation-821105340}) and (\ref{uz}) we have

\begin{eqnarray}
 l^2 & = &  - \displaystyle{\frac{{mv^{\,\prime} \left( {z_0 } \right)}}{{z_0 }}} =  - \frac{m}{{z_0 }}\left[ { - \frac{1}{2}\frac{{a\left( {4z_0^3  + 2bz_0 } \right)}}{{b^2 \sqrt {z_0^4  + bz_0^2 } } }- 2c} \right] \nonumber \\ 
 &  = & m\left[  \frac{ a\left( 2z_0^2  + b \right) } {b^2 z_0 \sqrt {z_0^2  + b} }   - 2c \right] .
\end{eqnarray}
Taking this result into (\ref{apsidal angle}) we obtain



\begin{equation}
\Delta \theta _a  = \frac{\pi }{{\sqrt a }}\sqrt {\left[ {\left( {z_0^2  + b} \right)\left( { {\frac{{a\left( {2z_0^2  + b} \right)}}{{b^2 }} - 2cz_0 \sqrt {z_0^2  + b} } } \right)} \right]}  .
\end{equation}
The potentials given by (\ref{sol2})  generate orbits for which the apsidal angles depend only on the angular momentum.  In this third case it is not possible to find constants $b$ and $c$ in such a way that those angles do not depend also on the angular momentum.

\section{Final Remarks}
In this work we have introduced a novel method for the determination of the apsidal angle in an arbitrary central field of force. For this reason equation (\ref{apsidal angle}) should be considered as the main result of this paper. From this equation we have discussed the conditions under which the apsidal angles remain independent of the energy and angular momentum and consequently lead to bounded and closed orbits, and we have found that only the Newtonian and the isotropic oscillator potentials present such behavior. Moreover, from equation (\ref{apsidal angle}) we have also calculated explicitly the value of their associated apsidal angles and have found the well-known results $\pi/2$ and $\pi$, respectively. We can see also that these two special fields are strongly determined by the dependence of the centrifugal potential on $1/r^2$.  As a consequence, we have re-obtained  Bertrand's theorem in an alternative but non-perturbative way. After more than one hundred years since its publication Bertrand's theorem still fascinates us \cite{berrondo}. Any serious student of classical mechanics sooner or later will become at least intrigued by it.  No wonder that alternative proofs some of them very interesting -- see for instance the phase space approach in \cite{quilantan} or the hamiltonian one in \cite{salasbrito} --  were offered in the past and nowadays.  We have also obtained a third potential which we can lay aside because it does not meet the conditions that lead to bounded closed orbits. 

The analytical function techniques applied to the problem of finding the only central fields
that allow an entire class of bounded, closed orbits with a minimum
number of restrictions lead in a concise, straightforward way
directly to the two allowed fields. It is also worth to mention that, as far as the present authors are aware of is one of the few examples of the usefulness of the B\"urmann-Lagrange theorem in its most general formulation. Equation (\ref{apsidal angle}) seems promising and its applications go beyond the rederivation of Bertrand's results. In fact, it can be taken as the starting point in the discussion of problems related to precession phenomena. Work by the present authors in this direction is in progress and results will be published elsewhere.

\newpage
%


\begin{figure*}
\includegraphics{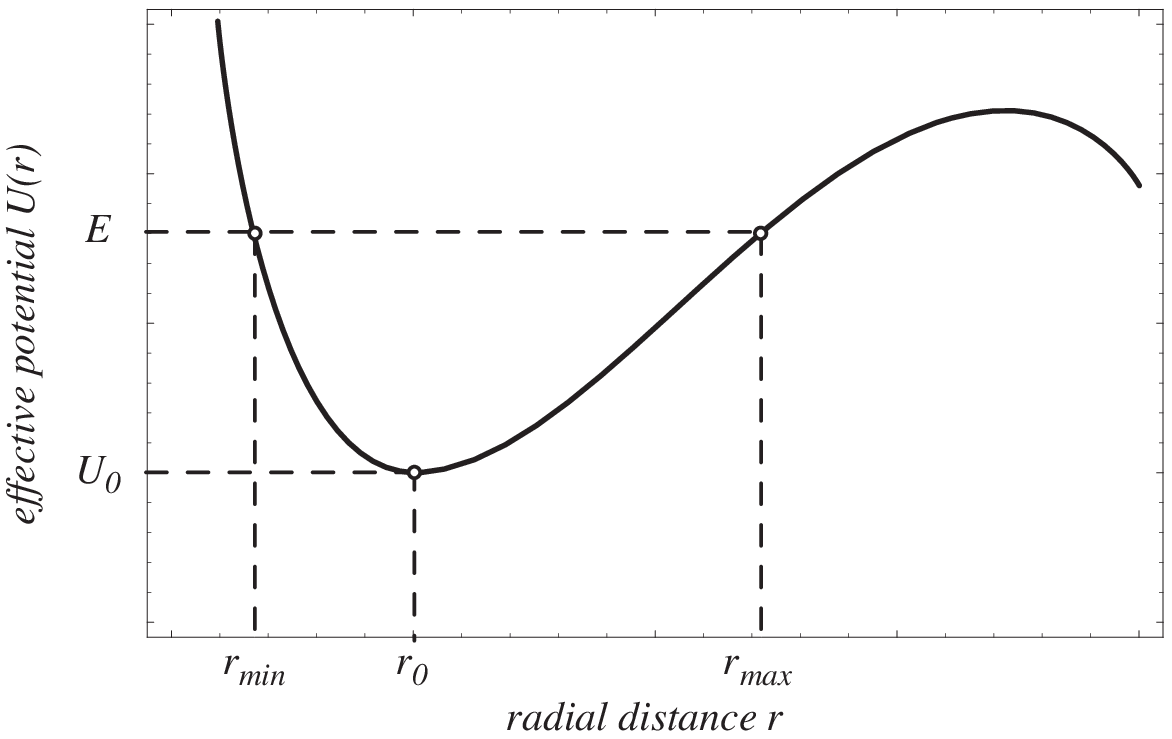}
\caption{General form of the effective potential energy.}
\label{Potential Plot}
\end{figure*}

\begin{figure*}
\begin{center}
\begin{pspicture}(-5,-5)(5,5)
\psset{arrowsize=0.1 3}%
\pscircle[linewidth=0.25mm](0,0){3}%
\psdot(0,2)%
\psdot(0,-2)%
\psarc(0,2){0.15}{-55}{230}%
\psarc(0,-2){0.15}{130}{415}%
\rput(0.5,2.0){$i\,b$}%
\rput(0.5,-2){$i\,b$}%
\rput(2,-0.3){$z_0$}%
\rput(-2.17,-0.3){$-z_0$}%
\rput(2,3){$C_R$}%
\rput(-0.65,2){$C_1$}%
\rput(-0.65,-2){$C_2$}%
\psline[linewidth=0.30mm]{-}(-0.1,-1.88)(-0.1,1.88) %
\psline[linewidth=0.30mm]{-}(0.1,-1.88)(0.1, 1.88)%
\psline[linewidth=0.15mm]{->}(-4,0)(4,0)%
\psline[linewidth=0.15mm]{->}(0,-4)(0,4)%
\psdot(-2,0)%
\psdot(2,0)%
\end{pspicture}
\caption{Contour}%
\label{contour}
\end{center}
\end{figure*}
%


\end{document}